\documentclass[11pt, a4paper]{article}
\usepackage[utf8]{inputenc}
\usepackage{amsmath, amssymb, amsfonts, bm, cite, float}
\usepackage{geometry}
\begin{document}

\begin{titlepage}

\vspace{2cm}

\begin{center}
{\Large\bf Off-Shell Supersymmetry Algebra \\ in the Lorentzian IIB Matrix Model:\\ Algebraic Constraints and a $\kappa$-Minkowski-Like Sector}
\vspace{2cm}

{\large Tetsuyuki Muramatsu\footnote{The views expressed in this paper are solely those of the author and do not necessarily reflect those of the affiliated organization.}} \\

\vspace{1cm}
{\it Tokai National Higher Education and Research System \\
Furo-cho, Chikusa-ku, Nagoya, 464-8601, Japan} \\
\vspace{2cm}

{\bf Abstract}
\end{center}
The Lorentzian IIB matrix model provides a non-perturbative framework for
emergent spacetime from matrix degrees of freedom. We study whether algebraic
consistency constrains such structures by imposing restricted off-shell
supersymmetry closure, modulo gauge transformations and explicitly identified
trivial symmetries, on a CPT-even low-order effective-action ansatz with
anisotropic background fields, without imposing their equations of motion.

The zeroth-order Ward identity forces the scalar ansatz to be constant. At
order two, retaining distinct macroscopic and internal transformation
normalizations leads to a closure-compatible block-diagonal branch. On the
nontrivial internal branch, Clifford-algebra identities force the internal
non-Abelian flux to vanish.

In four dimensions, a distinct dual-flux remainder can be absorbed into a
Lorentz-type rotation when the macroscopic matrices form a non-degenerate
coordinate sector. Within a linear absorption ansatz, the rank-three
coefficient tensor is the Hodge dual of a vector. Macroscopic spatial isotropy
selects its timelike orientation, yielding a $\kappa$-Minkowski-like algebra.
Finite-dimensional Hermitian representations make the spatial sector trivial,
so a nontrivial realization requires an infinite-dimensional limit with
unbounded coordinate operators. The different weights of the spatial and
internal generators under the adjoint action of $B_0$ are purely algebraic and
are not interpreted as physical time evolution or dynamical compactification.

\end{titlepage}

\newpage

\section{Introduction}
\label{sec:introduction}

The IIB matrix model (the IKKT model) \cite{IKKT_1997} was proposed as a
non-perturbative formulation of superstring theory, in which macroscopic
spacetime and gravity may emerge from microscopic $N\times N$ matrix degrees
of freedom. Its Lorentzian version incorporates a distinguished temporal
direction and provides a setting for studying emergent spacetime.
Non-perturbative studies, including Monte Carlo simulations and analyses of
classical equations of motion, have shown that the original nine-dimensional
spatial symmetry can be dynamically broken, leading to an expanding
three-dimensional macroscopic space
\cite{Kim_Nishimura_Tsuchiya_2012, Ito_etal_2014, Nishimura_Tsuchiya_2019}.
These dynamical approaches are complemented by theoretical studies of
emergent spacetime, matrix-model cosmology, and noncommutative geometries
\cite{Tsuchiya_Asano_2020, Steinacker_2010, Majid_Ruegg_1994,
Lukierski_etal_1991, Noncommutative_Geometry_1}.

Candidate spacetime structures, including $\kappa$-Minkowski-type
noncommutative algebras, have often been explored in connection with specific
classical or dynamical background configurations. As a complementary approach,
the present paper investigates whether such structures can also be constrained
by algebraic consistency conditions. Motivated by supersymmetry Ward identities
and algebraic closure constraints familiar from matrix models and related
supersymmetric systems \cite{BFSS_1997, Kazama_Muramatsu}, we address this
problem from the viewpoint of supersymmetry closure. To isolate the algebraic
constraints in a controlled setting, the analysis is performed within a local,
finite-order, CPT-even effective action ansatz in Minkowski signature. We
study the constraints imposed on background fields by requiring the
supersymmetry transformations to close. The background equations of motion are 
never set to zero. When equation-of-motion-proportional terms are mentioned 
below, they are retained explicitly as trivial-symmetry structures; the 
restricted closure test is understood modulo gauge transformations and explicitly 
identified trivial symmetries. Throughout this paper, ``off-shell closure'' is 
used strictly in this restricted sense; it does not refer to a complete 
auxiliary-field off-shell formulation of ten-dimensional super-Yang--Mills theory.

In our previous work \cite{Muramatsu_2026}, a restricted closure analysis
applied to isotropic backgrounds led to a non-renormalization theorem,
while a nontrivial escape route algebraically selected a Euclidean
(anti-)self-dual sector.

The present paper starts from the same obstruction mechanism but
investigates a different branch of the closure problem. The present paper does 
not attempt to absorb the ten-dimensional five-form obstruction itself. Rather, 
after passing to the anisotropic block-diagonal $SO(1,3)\times SO(6)$ branch, 
we analyze whether the distinct four-dimensional dual-flux remainder arising 
in the anisotropic analysis can be represented as a Lorentz-type rotation of 
the macroscopic matrices.

For the convenience of the reader, the obstruction terms and definitions required
for the present discussion are briefly recalled in Sec.~2 before
turning to the anisotropic Lorentzian analysis.

The paper is organized as follows. Section~2 introduces the anisotropic
decomposition of the IIB matrix model, the order-0 term, and the relation to our previous work. Section~3 derives
the block-diagonal sector and the vanishing of the internal flux from closure
constraints. Section~4 analyzes Lorentz-type absorption and derives a
$\kappa$-Minkowski-like algebra. Section~5 discusses the macroscopic time
direction, the finite-dimensional obstruction, and the adjoint weights. Section~6 gives the conclusions.
Technical details are collected in Appendices A--C.

\section{IIB Matrix Model and Anisotropic Decomposition}
\label{sec:model_and_decomposition}

\subsection{The Matrix Model and the Order Expansion Scheme}

Our starting point is the IIB matrix model \cite{IKKT_1997}. The degrees of freedom are ten $N \times N$ traceless Hermitian matrices $X_M$ and a $(1+9)$-dimensional Majorana-Weyl spinor $\Psi$. The classical action is
\begin{equation}
    S = -\frac{1}{g^2} \mathrm{Tr} \left( \frac{1}{4} [X_M, X_L][X^M, X^L] + \frac{1}{2} \bar{\Psi} \Gamma^M [X_M, \Psi] \right) .
\end{equation}
Vector indices $M,L$ are contracted with the Minkowski metric $\eta_{ML}$. Explicit conventions for the metric and the spinor decomposition are summarized in Appendix~\ref{app:conventions}.

We decompose the variables into classical background fields and quantum
fluctuations, $X_M=B_M+A_M$ and $\Psi=\psi+\tilde{\Psi}$, and denote the
background fermion by $\psi$. To systematically construct a derivative
expansion in the effective theory, we follow the methodology established in
our previous work \cite{Muramatsu_2026} and adopt an order expansion scheme in
which $\mathcal{O}(1)$ is assigned to non-Abelian commutators $[B_M,\cdot]$
and to background fermion bilinears, whereas the bare background field $B_M$
itself is assigned $\mathcal{O}(0)$. Compared with the previous leading-order
analysis, we also keep a possible order-0 scalar term.

We next analyze a sector in which the $SO(1,9)$ Lorentz symmetry is separated
into $SO(1,3)\times SO(6)$. Accordingly, we decompose the bosonic background
field $B_M$ into four macroscopic spacetime components $B_\mu$ and six
internal components $\phi_a$. We define the normalized traces characterizing
the corresponding macroscopic scales as
\begin{equation}
    x^2 = \frac{1}{N} \mathrm{Tr}(B_\mu B^\mu) , \quad y^2 = \frac{1}{N} \mathrm{Tr}(\phi_a \phi_a) .
\end{equation}

\subsection{The Order-0 Term}

The truncation used below is defined by the discrete symmetries of the
ten-dimensional supersymmetric Yang--Mills theory before dimensional
reduction. In particular, we use the charge-conjugation symmetry $C$ of the
parent theory and its CPT invariance as selection rules for the reduced
effective action. After dimensional reduction to the zero-dimensional matrix
model, a literal separation into $P$ and $T$ transformations is no longer
meaningful in the usual space-time sense. The relevant sign/transposition
operation in the reduced matrix model should therefore be understood as the
descendant of the charge-conjugation symmetry of the ten-dimensional parent
theory, not as an independent parity symmetry of the zero-dimensional model.

With this convention, CPT invariance of the parent theory restricts the
minimal ansatz to the inherited CPT-even sector. In particular, the effective
action is expanded in even orders,
\[
\Gamma = \Gamma^{(0)} + \Gamma^{(2)} + \dots .
\]
Possible terms outside this inherited CPT-even sector, as well as additional
higher-derivative tensor structures, are not included in the minimal ansatz
considered here.

We first consider the lowest-order scalar term
\[
    \Gamma^{(0)}=\Gamma^{(0)}(x^2,y^2),
    \qquad
    x^2=\frac{1}{N}\mathrm{Tr}(B_\mu B^\mu),
    \quad
    y^2=\frac{1}{N}\mathrm{Tr}(\phi_a\phi_a).
\]
In the restricted off-shell sense described above, the supersymmetry Ward
identity requires its variation under the leading transformation to vanish
identically,
\[
    \delta_\epsilon^{(1)}\Gamma^{(0)}=0 .
\]
Applying the chain rule and multiplying the resulting Ward identity by the common nonzero factor $N$, we obtain
\begin{equation}
    2c_1 \frac{\partial \Gamma^{(0)}}{\partial x^2}
    \mathrm{Tr}\!\left(B_\mu \bar{\epsilon}_I\gamma^\mu\psi^I\right)
    +
    2c_2 \frac{\partial \Gamma^{(0)}}{\partial y^2}
    \mathrm{Tr}\!\left(
    \phi_a \bar{\epsilon}_I(\gamma_5\rho^a)^I{}_{J}\psi^J
    \right)
    =0 .
\end{equation}
On the nontrivial branch ($c_1 c_2 \neq 0$), and for generic backgrounds for which the two trace structures are independent, the Ward identity implies
\begin{equation}
    \frac{\partial \Gamma^{(0)}}{\partial x^2}
    =
    \frac{\partial \Gamma^{(0)}}{\partial y^2}
    =0 .
\end{equation}
Therefore $\Gamma^{(0)}$ is restricted to a constant within the present
order-0 ansatz. Its value is not fixed by the restricted closure analysis
and is left as an undetermined non-perturbative contribution. The following
analysis focuses instead on the local algebraic constraints at order 2.

\subsection{Order-2 Effective Operators and Transformations}

The bosonic terms at $\mathcal{O}(2)$ can be constructed from quadratic forms
of the non-Abelian fluxes $\mathcal{F}_{ML}=i[B_M,B_L]$. Single-trace terms
linear in a commutator vanish because
$\mathrm{Tr}(\mathcal{F}_{ML})=0$. Possible dependence on lower-order multi-trace invariants, including those
built from $x^2$ and $y^2$, is absorbed into the coefficient functions
$f_i(x^2,y^2)$.

The following minimal set of order-2 operators is sufficient for the closure
channel analyzed below:
\begin{align}
    \Gamma^{(2)} &= \mathrm{Tr} \Big[ f_1(x^2, y^2) \mathcal{F}_{\mu\nu}\mathcal{F}^{\mu\nu} + f_2(x^2, y^2) (2\mathcal{F}_{\mu a}\mathcal{F}^{\mu a}) + f_3(x^2, y^2) \mathcal{F}_{ab}\mathcal{F}^{ab} \nonumber \\
    &\quad + g_1(x^2, y^2) \bar{\psi}_I \gamma^\mu [B_\mu, \psi^I] + g_2(x^2, y^2) \bar{\psi}_I (\gamma_5 \rho^a)^I{}_J [\phi_a, \psi^J] + \dots \Big] ,
\end{align}
where $\mathcal{F}_{\mu a}=i[B_\mu,\phi_a]$ is the cross-flux. The
ellipsis includes four-fermion terms and additional allowed tensor structures,
which are not needed for the purely bosonic closure channel analyzed below.

We parameterize the order-1 transformations of the fields as
\begin{equation}
    \delta_\epsilon^{(1)} B_\mu = c_1(x^2, y^2) \bar{\epsilon}_I \gamma_\mu \psi^I , \quad \delta_\epsilon^{(1)} \phi_a = c_2(x^2, y^2) \bar{\epsilon}_I (\gamma_5 \rho_a)^I{}_J \psi^J . \label{eq:boson_trans}
\end{equation}
The corresponding generic fermion transformation is parameterized as
\begin{align}
    \delta_{\epsilon}^{(1)} \psi^{I} &= w_1 \mathcal{F}_{\mu\nu} (\gamma^{\mu\nu} \epsilon)^{I} + w_2 \mathcal{F}_{ab} (\rho^{ab} \epsilon)^{I} + 2w_3 \mathcal{F}_{\mu a} (\gamma^{\mu} \gamma_5 \rho^{a} \epsilon)^{I} + \dots \label{eq:fermion_trans}
\end{align}

\subsection{Relation to the isotropic obstruction of Ref.~\cite{Muramatsu_2026}}
For completeness, we briefly recall the part of the isotropic closure analysis of Ref.~\cite{Muramatsu_2026} that is needed below.
In that work, an isotropic radial variable $r$ was defined as
\begin{equation}
    r^2 = \frac{1}{N}\mathrm{Tr}(B_M B^M) .
\end{equation}
Here $\lambda$ denotes the background fermion in the notation of Ref.~\cite{Muramatsu_2026}, corresponding to $\psi$ in the notation used in the present paper.

In the isotropic analysis, the leading bosonic transformation was
\begin{equation}
    \delta_\epsilon^{(1)} B_M = c(r) \bar{\epsilon} \Gamma_M \lambda .
\end{equation}
It follows that the induced variation of the radial variable is
\begin{equation}
\begin{split}
    \delta_\epsilon^{(1)} r &= \frac{1}{2rN} \mathrm{Tr}\!\left[ (\delta_\epsilon^{(1)} B_M) B^M + B_M (\delta_\epsilon^{(1)} B^M) \right] \\
    &= \frac{c(r)}{rN} \mathrm{Tr}\!\left( B^M \bar{\epsilon} \Gamma_M \lambda \right).
\end{split}
\end{equation}
Therefore, the variation of the field-dependent coefficient is
\begin{equation}
    \delta_\epsilon^{(1)} c(r) = c'(r) \delta_\epsilon^{(1)} r .
\end{equation}

Substitution of this variation into the commutator generates derivative contributions of the form
\begin{equation}
\Delta B^{(\mathrm{deriv})}_{M}
=
\frac{c(r)c'(r)}{rN}
\Bigl[
{\rm Tr}
\bigl(
B^{L}\bar\epsilon_{1}\Gamma_{L}\lambda
\bigr)
\bar\epsilon_{2}\Gamma_{M}\lambda
-
(1\leftrightarrow2)
\Bigr].
\label{eq:ref12-review-1}
\end{equation}

Defining the antisymmetric spinor bilinears
\begin{equation}
    v^{(n)}_{M_1\dots M_n} := \bar{\epsilon}_2 \Gamma_{M_1\dots M_n} \epsilon_1 ,
\end{equation}
the relevant antisymmetrized Fierz decomposition for the ten-dimensional Majorana-Weyl parameters is given by
\begin{equation}
    \epsilon_1 \bar{\epsilon}_2 - \epsilon_2 \bar{\epsilon}_1 = \frac{1}{16} v^{(1)}_M \Gamma^M + \frac{1}{3840} v^{(5)}_{M_1\dots M_5} \Gamma^{M_1\dots M_5} .
\end{equation}

After this Fierz rearrangement, the derivative terms were shown to contain an irreducible five-form contribution,
\begin{equation}
\Delta B^{(\mathrm{deriv})}_{M}
\supset
\frac{c(r)c'(r)}{rN}
v^{(5)}_{L_1 \dots L_5}
{\rm Tr}
\!\left(
B^{L}\bar\lambda
\Gamma_{L}
\Gamma^{L_1 \dots L_5}
\Gamma_{M}
\lambda
\right)\,.
\label{eq:ref12-review-2}
\end{equation}
This term is not matched by any of the gauge, Lorentz-type, or equation-of-motion structures allowed in the closure ansatz. In particular, the $\mathrm{SU}(N)$ gauge algebra contains no generator carrying the independent five-form parameter.
Consequently, closure forces $c(r)c'(r)=0$. On the nontrivial branch $c(r)\neq 0$, this gives $c'(r)=0$, which is the origin of the non-renormalization result in Ref.~\cite{Muramatsu_2026}.

The ten-dimensional five-form obstruction recalled above is not the term absorbed in the present analysis. After restricting to the anisotropic block-diagonal branch, the four-dimensional commutator instead produces a distinct dual-flux remainder $v^\nu\tilde{\mathcal{F}}_{\nu\mu}$, derived explicitly in Appendix~\ref{app:lorentz_calculation}. Section~\ref{sec:lorentz_absorption} asks whether this distinct remainder can be represented as a Lorentz-type rotation of the macroscopic matrices.

\section{Closure Constraints and the Block-Diagonal Sector}
\label{sec:closure_and_block_diagonal}

\subsection{Constraints from Single Gauge Invariance}

In the present truncation, consistency requires the commutator
$[\delta_{\epsilon_1}^{(1)},\delta_{\epsilon_2}^{(1)}]$ 
to reproduce a gauge transformation with a parameter proportional to
\(v^\mu B_\mu+v^a\phi_a\).

The coefficients $v^\mu$ and $v^a$
are the corresponding spinor-bilinear parameters, and their relative
normalization is fixed by this single gauge parameter.

We restrict to the generic cross-coupled branch $w_3\neq0$. The exceptional branch $w_3=0$, in which the cross-flux does not enter the fermion transformation, requires a separate analysis and is not considered here.

As derived in Appendix~\ref{app:coefficient_matching}, with the definitions of $v^\mu$ and $v^a$ fixed, the vector parts of the two commutators can be summarized with a common overall factor $\mathcal{N} = -2$ as
\begin{align}
\left.[\delta_{\epsilon_1}^{(1)},\delta_{\epsilon_2}^{(1)}]B_\mu\right|_{\mathcal F_{\rho\sigma}}
&=
-2c_1w_1v^\nu\mathcal F_{\nu\mu},
\\
\left.[\delta_{\epsilon_1}^{(1)},\delta_{\epsilon_2}^{(1)}]B_\mu\right|_{\mathcal F_{\rho a}}
&=
-2c_1w_3v^a\mathcal F_{a\mu},
\\
\left.[\delta_{\epsilon_1}^{(1)},\delta_{\epsilon_2}^{(1)}]\phi_a\right|_{\mathcal F_{\mu b}}
&=
-2c_2w_3v^\mu\mathcal F_{\mu a},
\\
\left.[\delta_{\epsilon_1}^{(1)},\delta_{\epsilon_2}^{(1)}]\phi_a\right|_{\mathcal F_{bc}}
&=
-2c_2w_2v^b\mathcal F_{ba}.
\end{align}
Matching these four structures to a single gauge transformation with
a parameter proportional to
\(v^\mu B_\mu+v^a\phi_a\)
therefore requires
\begin{equation}
    c_1 w_1 = c_1 w_3 = c_2 w_3 = c_2 w_2 \equiv \kappa_{\mathrm{cl}} .
\end{equation}

On the generic cross-coupled branch $w_3\neq0$, the algebraic relation $c_1w_3=c_2w_3$ implies
\begin{equation}
    c_1(x^2, y^2) = c_2(x^2, y^2) \equiv c(x^2, y^2) .
\end{equation}
Thus, the cross-flux generically ties the transformation coefficients.

\subsection{Derivative Terms and Anisotropic Splitting}

Proceeding with $c_1=c_2\equiv c$, the variation of the common coefficient is
\begin{equation}
    \delta_\epsilon^{(1)} c = \frac{2c}{N} \left[ \frac{\partial c}{\partial x^2} \mathrm{Tr}(B_\nu \bar{\epsilon}_I \gamma^\nu \psi^I) + \frac{\partial c}{\partial y^2} \mathrm{Tr}(\phi_a \bar{\epsilon}_I (\gamma_5 \rho^a)^I{}_J \psi^J) \right] .
\end{equation}
This variation generates derivative terms in the commutator of the form
\begin{equation}
\begin{split}
    [\delta_{\epsilon_1}^{(1)}, \delta_{\epsilon_2}^{(1)}] B_{\mu} \supset \frac{2c}{N} \Bigg[ & \frac{\partial c}{\partial x^2} \mathrm{Tr} \left( B_\nu \bar{\epsilon}_{1I} \gamma^\nu \psi^I \right) \\
    &+ \frac{\partial c}{\partial y^2} \mathrm{Tr} \left( \phi_a \bar{\epsilon}_{1I} (\gamma_5 \rho^a)^I{}_J \psi^J \right) \Bigg] \bar{\epsilon}_{2K} \gamma_{\mu} \psi^{K} - (1 \leftrightarrow 2) .
\end{split}
\end{equation}
The two derivative terms contain independent $SO(1,3)$ and $SO(6)$ trace structures. Since neither structure is reproduced by the allowed local closure generators, their coefficients must vanish separately.

The obstruction is not merely the presence of a bare $B_\nu$, but the
matrix-valued Grassmann-even structure of the form $\mathrm{Tr}(B\psi)\psi$.
This structure is not reproduced by the standard bosonic gauge parameter or
by a linear Lorentz-type rotation. Nor can it be cancelled by the trivial terms allowed in this truncation,
namely terms proportional to the leading bosonic equations of motion, since
those terms do not have the required $\mathrm{Tr}(B\psi)\psi$ matrix
structure.

Apart from absorptions involving non-local operators, infinite-order field
redefinitions, or composite operators tuned to cancel this structure, this
mismatch cannot be resolved within the present class of local redefinitions.
The requirement of algebraic closure therefore restricts the coefficient to a
constant, $\partial c/\partial x^2 = \partial c/\partial y^2 = 0$.

The preceding analysis shows that, on the cross-coupled branch
$\mathcal{F}_{\mu a}\neq0$ with $w_3\neq0$, matching to a single
gauge parameter enforces $c_1=c_2$. The derivative terms then require
this common coefficient to be constant.

\subsection{Block-Diagonal Sector and Exact Vanishing of Internal Flux}

To retain an anisotropic splitting between the macroscopic and
internal supersymmetry normalizations, $c_1\neq c_2$, we therefore
consider the closure-compatible block-diagonal branch
\begin{equation}
    [B_\mu,\phi_a]=0.
    \label{eq:block_diagonal_branch}
\end{equation}
Under this condition, the cross-flux vanishes identically:
\begin{equation}
    \mathcal{F}_{\mu a}
    =
    i[B_\mu,\phi_a]
    =
    0.
\end{equation}
Consequently, the cross-sector matching condition no longer enforces
$c_1=c_2$. The two coefficients may therefore be treated
independently at the level of the cross-sector algebra.

The block-diagonal condition does not, however, remove the variations
of the coefficient functions $c_i(x^2,y^2)$. From
\begin{equation}
    x^2
    =
    \frac{1}{N}\mathrm{Tr}(B_\mu B^\mu),
    \qquad
    y^2
    =
    \frac{1}{N}\mathrm{Tr}(\phi_a\phi_a),
\end{equation}
and the leading bosonic transformations, one obtains
\begin{align}
    \delta_\epsilon^{(1)}x^2
    &=
    \frac{2c_1}{N}
    \mathrm{Tr}
    \left(
        B_\mu
        \bar{\epsilon}_I\gamma^\mu\psi^I
    \right),
    \label{eq:block_variation_x2}
    \\
    \delta_\epsilon^{(1)}y^2
    &=
    \frac{2c_2}{N}
    \mathrm{Tr}
    \left(
        \phi_a
        \bar{\epsilon}_I
        (\gamma_5\rho^a)^I{}_{J}
        \psi^J
    \right).
    \label{eq:block_variation_y2}
\end{align}
Therefore, for $i=1,2$, the chain rule gives
\begin{equation}
\begin{split}
    \delta_\epsilon^{(1)}c_i
    =
    \frac{2}{N}
    \Bigg[
        &c_1
        \frac{\partial c_i}{\partial x^2}
        \mathrm{Tr}
        \left(
            B_\mu
            \bar{\epsilon}_I\gamma^\mu\psi^I
        \right)
        \\
        &+
        c_2
        \frac{\partial c_i}{\partial y^2}
        \mathrm{Tr}
        \left(
            \phi_a
            \bar{\epsilon}_I
            (\gamma_5\rho^a)^I{}_{J}
            \psi^J
        \right)
    \Bigg].
\end{split}
\label{eq:block_variation_ci}
\end{equation}

For the macroscopic matrix, the variation of $c_1$ produces the
derivative contribution
\begin{equation}
\begin{split}
    \left.
    [\delta_{\epsilon_1}^{(1)},
     \delta_{\epsilon_2}^{(1)}]
    B_\mu
    \right|_{\mathrm{deriv}}
    =
    \frac{2}{N}
    \Bigg[
        &c_1
        \frac{\partial c_1}{\partial x^2}
        \mathrm{Tr}
        \left(
            B_\nu
            \bar{\epsilon}_{1I}\gamma^\nu\psi^I
        \right)
        \\
        &+
        c_2
        \frac{\partial c_1}{\partial y^2}
        \mathrm{Tr}
        \left(
            \phi_a
            \bar{\epsilon}_{1I}
            (\gamma_5\rho^a)^I{}_{J}
            \psi^J
        \right)
    \Bigg]
    \bar{\epsilon}_{2K}\gamma_\mu\psi^K
    -
    (1\leftrightarrow2).
\end{split}
\label{eq:block_derivative_commutator_B}
\end{equation}
Similarly, the variation of $c_2$ produces
\begin{equation}
\begin{split}
    \left.
    [\delta_{\epsilon_1}^{(1)},
     \delta_{\epsilon_2}^{(1)}]
    \phi_a
    \right|_{\mathrm{deriv}}
    =
    \frac{2}{N}
    \Bigg[
        &c_1
        \frac{\partial c_2}{\partial x^2}
        \mathrm{Tr}
        \left(
            B_\nu
            \bar{\epsilon}_{1I}\gamma^\nu\psi^I
        \right)
        \\
        &+
        c_2
        \frac{\partial c_2}{\partial y^2}
        \mathrm{Tr}
        \left(
            \phi_b
            \bar{\epsilon}_{1I}
            (\gamma_5\rho^b)^I{}_{J}
            \psi^J
        \right)
    \Bigg]
    \bar{\epsilon}_{2K}
    (\gamma_5\rho_a)^K{}_{L}
    \psi^L
    -
    (1\leftrightarrow2).
\end{split}
\label{eq:block_derivative_commutator_phi}
\end{equation}

The two traces in
Eqs.~\eqref{eq:block_derivative_commutator_B} and
\eqref{eq:block_derivative_commutator_phi} are independent
$SO(1,3)$ and $SO(6)$ structures. They generate the same type of
matrix-valued Grassmann-even terms,
$\mathrm{Tr}(B\psi)\psi$ and
$\mathrm{Tr}(\phi\psi)\psi$, as in the cross-coupled analysis above.
Neither structure is reproduced by the gauge, Lorentz-type, or
equation-of-motion-proportional trivial transformations allowed in
the present local truncation.

Closure for arbitrary supersymmetry parameters therefore requires
\begin{equation}
\begin{aligned}
    c_1\frac{\partial c_1}{\partial x^2}
    &=
    0,
    &
    c_2\frac{\partial c_1}{\partial y^2}
    &=
    0,
    \\
    c_1\frac{\partial c_2}{\partial x^2}
    &=
    0,
    &
    c_2\frac{\partial c_2}{\partial y^2}
    &=
    0.
\end{aligned}
\label{eq:block_derivative_constraints}
\end{equation}
On the nontrivial branch
\begin{equation}
    c_1c_2\neq0,
\end{equation}
these conditions reduce to
\begin{equation}
    \frac{\partial c_i}{\partial x^2}
    =
    \frac{\partial c_i}{\partial y^2}
    =
    0,
    \qquad
    i=1,2.
    \label{eq:block_constant_coefficients}
\end{equation}
Thus, both $c_1$ and $c_2$ are constants within the present local
finite-order ansatz. The block-diagonal branch permits an anisotropic
splitting
\begin{equation}
    c_1\neq c_2
\end{equation}
between two constant supersymmetry normalizations; it does not permit
a nonconstant dependence of either coefficient on $x^2$ or $y^2$.

The remaining intra-sector closure conditions continue to constrain
the allowed flux and coordinate-algebra sectors. 
The internal condition is analyzed
below, while the four-dimensional condition is considered in
Section~\ref{sec:lorentz_absorption}.

In the six-dimensional internal sector, on the nontrivial internal
branch
\begin{equation}
    c_2w_2\neq0,
\end{equation}
the internal commutator contains the remainder
\begin{equation}
\begin{split}
    \Delta\phi_a^{(\mathrm{rem})}
    =
    c_2w_2
    \Bigl[
        \bar{\epsilon}_2
        \gamma_5\rho_{abc}
        \epsilon_1
        -
        (1\leftrightarrow2)
    \Bigr]
    \mathcal{F}^{bc}.
\end{split}
\label{eq:internal_clifford_remainder}
\end{equation}

Using the fixed-chirality condition
\eqref{eq:fixed_four_dimensional_chirality} and the fact that
$\gamma_5$ commutes with the internal Clifford matrices, we have
\begin{equation}
    \bar{\epsilon}_2
    \gamma_5\rho_{abc}
    \epsilon_1
    =
    \chi\,
    \bar{\epsilon}_2
    \rho_{abc}
    \epsilon_1 .
\end{equation}
Equation~\eqref{eq:internal_clifford_remainder} can therefore be
written as
\begin{equation}
\begin{split}
    \Delta\phi_a^{(\mathrm{rem})}
    =
    \chi c_2w_2
    \Bigl[
        \bar{\epsilon}_2
        \rho_{abc}
        \epsilon_1
        -
        (1\leftrightarrow2)
    \Bigr]
    \mathcal{F}^{bc}.
\end{split}
\label{eq:internal_clifford_remainder_reduced}
\end{equation}

Since $\chi=\pm1$ and $c_2w_2\neq0$, closure for arbitrary
supersymmetry parameters requires the coefficient of this independent
spinor-bilinear structure to vanish. By non-degeneracy of the spinor
pairing in this Clifford channel, this gives
\begin{equation}
    \rho_{abc}\mathcal{F}^{bc}=0
\end{equation}
as an identity on the tensor product of the internal spinor space and
the matrix-valued color space.

The vanishing of the internal flux now follows from the internal
Clifford algebra. Multiplying the preceding identity by $\rho^a$ and
using
\begin{equation}
    \rho^a\rho_{abc}
    =
    4\rho_{bc}
\end{equation}
in six dimensions, we obtain
\begin{equation}
    4\rho_{bc}\mathcal{F}^{bc}=0.
\end{equation}
Because $\mathcal{F}^{bc}$ is matrix-valued in the color space, this
equation is understood as an identity on the tensor product of the
internal spinor space and the matrix space.

The fifteen matrices $\rho_{bc}$ are linearly independent in the
internal spinor space. Equivalently, their trace pairing is
non-degenerate:
\begin{equation}
    \mathrm{tr}_{\mathrm{sp}}
    \left(
        \rho^{de}\rho_{bc}
    \right)
    =
    C_{\rho}\,\delta^{de}_{bc},
    \qquad
    C_{\rho}\neq0.
\end{equation}
Applying this projection to
$4\rho_{bc}\mathcal{F}^{bc}=0$ shows that every matrix coefficient
vanishes, $\mathcal{F}^{de}=0$. Thus, on the nontrivial
internal closure branch $c_2w_2\neq0$, the internal non-Abelian flux
vanishes:
\begin{equation}
    \mathcal{F}_{ab}
    =
    i[\phi_a,\phi_b]
    =
    0.
\end{equation}

At this stage, this result should be understood purely as an
algebraic decoupling condition between the four-dimensional and
internal matrix sectors.

\section{Linear Lorentz Absorption and Spacetime Algebra}
\label{sec:lorentz_absorption}

In the four-dimensional spacetime sector, the closure generates the
distinct dual-flux remainder derived in
Eq.~\eqref{eq:four_dimensional_dual_remainder}:
\begin{equation}
    \Delta B_\mu^{(\mathrm{rem})}
    =
    \mathcal{C}_4\,
    v^\nu\widetilde{\mathcal{F}}_{\nu\mu},
    \qquad
    \mathcal{C}_4
    :=
    2i\chi c_1w_1.
    \label{eq:four_dimensional_remainder_main}
\end{equation}
We restrict below to the nontrivial four-dimensional branch
$\mathcal{C}_4\neq0$.

\subsection{Linear Lorentz-Absorption Ansatz}

We ask whether this remainder can be represented as a Lorentz-type
rotation of the emergent four-dimensional matrices,
\begin{equation}
    \delta_\omega B_\mu
    =
    \omega_{\mu\lambda}B^\lambda.
\end{equation}
This is only an algebraic identification inside the emergent
four-dimensional sector; it does not introduce a fundamental local
Lorentz gauge symmetry.

We restrict to the simplest ansatz in which the Lorentz parameter is
linear in the translation vector $v^\nu$:
\begin{equation}
    \omega_{\mu\lambda}
    =
    M_{\nu\mu\lambda}v^\nu.
\end{equation}
The Lorentz-type variation is then
\begin{equation}
    \delta_\omega B_\mu
    =
    v^\nu M_{\nu\mu\lambda}B^\lambda.
\end{equation}
Requiring exact absorption of
Eq.~\eqref{eq:four_dimensional_remainder_main} for arbitrary $v^\nu$
gives
\begin{equation}
    \mathcal{C}_4\,
    \widetilde{\mathcal{F}}_{\nu\mu}
    =
    M_{\nu\mu\lambda}B^\lambda.
    \label{eq:exact_linear_absorption}
\end{equation}

Since $\widetilde{\mathcal{F}}_{\nu\mu}$ is antisymmetric in $\nu$
and $\mu$, Eq.~\eqref{eq:exact_linear_absorption} implies
\begin{equation}
    \left(
        M_{\nu\mu\lambda}
        +
        M_{\mu\nu\lambda}
    \right)
    B^\lambda
    =
    0.
\end{equation}
We assume that the four macroscopic matrices are linearly independent
in the degree-one subspace of the coordinate algebra. Explicitly, for
scalar coefficients $a_\lambda$ allowed in the present ansatz,
\begin{equation}
    a_\lambda B^\lambda=0
    \qquad\Longrightarrow\qquad
    a_\lambda=0.
    \label{eq:degree_one_independence}
\end{equation}
This is the non-degeneracy assumption on the macroscopic coordinate
sector; representations with nontrivial degree-one relations among the
$B^\lambda$ lie outside the branch considered here. Applying
Eq.~\eqref{eq:degree_one_independence} gives
\begin{equation}
    M_{\nu\mu\lambda}
    =
    -M_{\mu\nu\lambda}.
\end{equation}

Furthermore, the Lorentz parameter satisfies
$\omega_{\mu\lambda}=-\omega_{\lambda\mu}$. Since
$\omega_{\mu\lambda}=M_{\nu\mu\lambda}v^\nu$ for arbitrary $v^\nu$,
we also have
\begin{equation}
    M_{\nu\mu\lambda}
    =
    -M_{\nu\lambda\mu}.
\end{equation}
These two antisymmetry conditions make $M_{\nu\mu\lambda}$ totally
antisymmetric. In four dimensions, it is therefore the Hodge dual of
a vector. Separating the direction of this vector from its overall
normalization, we parameterize the exact solution as
\begin{equation}
    M_{\nu\mu\lambda}
    =
    -\mathcal{C}_4\Theta\,
    \epsilon_{\sigma\nu\mu\lambda}m^\sigma.
    \label{eq:M_hodge_dual_exact}
\end{equation}

\subsection{Derivation of the Spacetime Algebra}

Substituting Eq.~\eqref{eq:M_hodge_dual_exact} into the exact absorption
condition and contracting the two Levi--Civita tensors gives, as shown in
Appendix~\ref{app:lorentz_calculation},
\begin{equation}
    \mathcal{F}^{\mu\nu}
    =
    -\Theta
    \left(
        m^\mu B^\nu
        -
        m^\nu B^\mu
    \right).
    \label{eq:exact_flux_relation}
\end{equation}
Using $\mathcal{F}^{\mu\nu}=i[B^\mu,B^\nu]$, we obtain
\begin{equation}
    [B^\mu,B^\nu]
    =
    i\Theta
    \left(
        m^\mu B^\nu
        -
        m^\nu B^\mu
    \right).
    \label{eq:spacetime_algebra}
\end{equation}
For Hermitian coordinate matrices and a real orientation vector $m^\mu$,
Hermiticity requires $\Theta$ to be real. For nonzero $m^\mu$,
macroscopic spatial isotropy requires $m^i=0$ and hence selects its
timelike orientation, yielding a $\kappa$-Minkowski-like algebra. Here
$\Theta$ corresponds, up to a sign convention, to the usual $1/\kappa$
parameter and controls the strength of the spacetime noncommutativity
\cite{Majid_Ruegg_1994,Lukierski_etal_1991}.

\section{Timelike $\kappa$-Minkowski Branch and Adjoint Weights}
\label{sec:emergent_time}

\subsection{Timelike Branch and the Adjoint Derivation Generated by $B_0$}

To make contact with the Lorentzian macroscopic setup, we impose macroscopic
spatial isotropy, namely the preservation of $SO(3)$ in the spatial directions.
Normalizing the timelike vector
to $m^\mu=(1,0,0,0)$ and lowering the temporal index ($\eta_{00}=-1$), we
obtain the time-space commutator\footnote{Choosing the orientation so that $\Theta>0$, the algebraic scaling weight of the spatial generators is positive.}:
\begin{align}
    [B^i, B^j] &= 0 , \label{eq:spatial_comm} \\
    [B_0, B_i] &= -i \Theta B_i . \label{eq:time_space_comm}
\end{align}

Once the timelike branch has been selected, $B_0$ provides a distinguished generator in the emergent coordinate algebra. The four-dimensional spatial matrices
and the internal matrices then obey
\begin{equation}
    [B_0,B_i]=-i\Theta B_i, \qquad [B_0,\phi_a]=0.
\end{equation}

\paragraph{Remark on background supersymmetry.}

The closure condition imposed above is logically distinct from the
supersymmetry invariance of a particular background. Closure
constrains the algebra of the transformations, whereas unbroken
supersymmetry requires the fermionic variation to vanish on the
background for at least one nonzero supersymmetry parameter.

For a purely bosonic background, this additional condition is
\begin{equation}
    \left.
    \delta_\epsilon^{(1)}\psi
    \right|_{\psi=0}
    =
    0 .
\end{equation}
In the block-diagonal branch,
\begin{equation}
    \mathcal{F}_{\mu a}=0,
    \qquad
    \mathcal{F}_{ab}=0,
\end{equation}
so that the leading fermionic variation reduces to
\begin{equation}
    w_1
    \mathcal{F}_{\mu\nu}
    \gamma^{\mu\nu}
    \epsilon
    =
    0 .
\end{equation}
On the nontrivial four-dimensional branch \(w_1\neq0\), background
supersymmetry therefore requires
\begin{equation}
    \mathcal{F}_{\mu\nu}
    \gamma^{\mu\nu}
    \epsilon
    =
    0
\label{eq:background_susy_condition}
\end{equation}
for some nonzero spinor \(\epsilon\).

For the timelike \(\kappa\)-Minkowski-like algebra derived above,
\begin{equation}
    \mathcal{F}_{0i}
    =
    i[B_0,B_i]
    =
    \Theta B_i ,
\end{equation}
which is nonzero in a nontrivial spatial representation. Thus,
Eq.~\eqref{eq:background_susy_condition} imposes a nontrivial
additional constraint and is not expected to admit a nonzero
solution for a generic representation.

The present closure analysis alone does not, however, exclude special
operator representations or special nonzero spinors for which
Eq.~\eqref{eq:background_susy_condition} is satisfied. We therefore
do not claim that every realization of the selected algebra
necessarily breaks supersymmetry. If the full nonperturbative matrix
dynamics selects a representation for which
Eq.~\eqref{eq:background_susy_condition} has no nonzero solution,
that realization would be interpreted as a spontaneously
supersymmetry-breaking background. Determining whether such solutions
exist requires a representation-specific analysis beyond the
restricted closure calculation performed here.

\subsection{The Finite-$N$ Obstruction and Internal Block Structure}

We now evaluate the derived commutation relations within a finite-dimensional
Hermitian matrix representation. Diagonalizing $B_0$ by a unitary
transformation and taking the $(m,n)$ matrix element yields, for the spatial
matrices,
\begin{equation}
    (t_m - t_n) (B_i)_{mn} = -i\Theta (B_i)_{mn} .
\end{equation}
Equivalently,
\begin{equation}
    \left(t_m-t_n+i\Theta\right)(B_i)_{mn}=0 .
\end{equation}
Since $B_0$ is Hermitian, $t_m-t_n$ is real. For real nonzero $\Theta$, the
coefficient $t_m-t_n+i\Theta$ is never zero, even in the degenerate case
$t_m=t_n$. Therefore $(B_i)_{mn}=0$ for all $m,n$, and each spatial matrix
vanishes.

This shows that the spacetime algebra has no nontrivial realization by
finite-dimensional Hermitian matrices. A nontrivial realization therefore
requires an infinite-dimensional limit involving unbounded coordinate operators. The
present paper fixes only the algebraic structure required by supersymmetry
closure; the corresponding double-scaling limit, including the scaling of
$g$, the normalization of observables, and the emergence of a finite stress
tensor, is left for future work.

Similarly, evaluating the block-diagonal condition $[B_0,\phi_a]=0$ in this
basis gives $(t_m-t_n)(\phi_a)_{mn}=0$. Thus $\phi_a$ has no off-diagonal
elements connecting different eigenspaces of $B_0$, while matrix elements
inside a degenerate block may remain. This yields a block-diagonal
structure $\phi_a=\bigoplus_t\phi_a(t)$ and eliminates mixing
between different eigenspaces of $B_0$.

\subsection{Adjoint Weights under $B_0$}

At the level of the abstract coordinate algebra, the commutation relations may be written as
\begin{equation}
    \operatorname{ad}_{B_0}(B_i) = [B_0, B_i] = -i\Theta B_i , \qquad \operatorname{ad}_{B_0}(\phi_a) = [B_0, \phi_a] = 0 .
\end{equation}
Thus, the spatial generators and the internal generators are eigenoperators of the derivation $\operatorname{ad}_{B_0}$, with different algebraic weights.

If a representation is chosen in which $B_0$ is self-adjoint and the exponentials and common invariant domains are well defined, one may formally exponentiate this derivation and write
\begin{equation}
    \alpha_s(\mathcal{O}) := e^{isB_0} \mathcal{O} e^{-isB_0} ,
\end{equation}
so that
\begin{equation}
    \alpha_s(B_i) = e^{\Theta s} B_i , \qquad \alpha_s(\phi_a) = \phi_a .
\end{equation}
These equations describe only the adjoint weights of the generators under the formal adjoint action of $B_0$. They are not equations of motion and do not determine the evolution of physical observables. The parameter $s$ is not a physical time coordinate, and no dynamical expansion or compactification is inferred from these relations.

Nevertheless, within the restricted closure ansatz, this distinction is a
nontrivial algebraic result: the closure conditions separate the spatial and
internal generators into different eigenspaces of
$\operatorname{ad}_{B_0}$ without invoking a Hamiltonian evolution
equation. It therefore provides an algebraic characterization of their
asymmetric embedding in the emergent coordinate algebra.

This should not be interpreted as a derivation of a dynamical compact internal
manifold. In the block-diagonal branch the internal matrices commute with the
macroscopic sector and satisfy $\mathcal{F}_{ab}=0$. Hence, in an irreducible
representation, they reduce to central moduli; nontrivial internal profiles
require a reducible, direct-integral, or otherwise continuum representation.

\section{Conclusion and Discussion}
\label{sec:conclusion_and_discussion}

In this paper, we have studied the algebraic constraints imposed by
supersymmetry closure on a CPT-even low-order effective-action ansatz for the
Lorentzian IIB matrix model in Minkowski signature, without using the
background equations of motion. The zeroth-order Ward identity forces the
scalar ansatz $\Gamma^{(0)}=\Gamma^{(0)}(x^2,y^2)$ to be constant. At order
two, closure constrains the effective transformation coefficients. Within the 
present local finite-order ansatz, retaining an anisotropic splitting between 
the macroscopic and internal transformation coefficients leads to a closure-compatible 
block-diagonal branch. On the nontrivial internal closure branch ($c_2w_2\neq0$), 
Clifford-algebra identities then force the internal non-Abelian flux to vanish, 
giving an algebraic decoupling of the internal sector.

In the four-dimensional sector, a distinct four-dimensional dual-flux remainder 
can be absorbed into a Lorentz-type rotation when the macroscopic matrices form 
a non-degenerate coordinate sector. Within the linear absorption ansatz, the 
rank-three coefficient tensor is constrained to be totally antisymmetric and is 
therefore the Hodge dual of a vector $m^\sigma$. Macroscopic spatial isotropy 
selects the timelike orientation of this vector, yielding a $\kappa$-Minkowski-like 
algebra. This algebra should be interpreted as an algebraically selected 
structure within the restricted class of effective descriptions considered in 
this work, rather than as a unique consequence of the full matrix model. The 
representation theory and continuum geometry of this $\kappa$-Minkowski-like 
sector remain important open questions.

The \(\kappa\)-Minkowski-like algebra obtained here does not
generically define a supersymmetry-preserving background. In a
nontrivial spatial representation it has
\(\mathcal{F}_{0i}=\Theta B_i\neq0\), so preservation of linearly
realized supersymmetry imposes the additional background-invariance
condition given in
Eq.~\eqref{eq:background_susy_condition}. The closure analysis by
itself does not determine whether this equation admits a nonzero
solution in a specific infinite-dimensional operator representation.
Special supersymmetry-preserving realizations are therefore not
excluded by the present argument. If a representation selected by
the full matrix dynamics admits no nonzero solution, it would instead
constitute a spontaneously supersymmetry-breaking phase.

In a formal unbounded-operator representation, the adjoint action of
\(B_0\) distinguishes the spatial and internal generators by
assigning them different algebraic weights. This distinction is
purely algebraic: without an independently derived Hamiltonian,
physical clock, state or measure, and continuum geometry, it does
not imply physical time evolution, spatial expansion, or dynamical
compactification. Within the restricted ansatz, however, the
nontrivial result is that the closure conditions themselves fix this
separation of adjoint weights, distinguishing the two sectors at the
level of the emergent coordinate algebra.

Finally, finite-dimensional Hermitian representations make the
spatial sector trivial, so a nontrivial realization requires an
infinite-dimensional limit involving unbounded coordinate operators.
The representation-specific supersymmetry condition and the
associated double-scaling or continuum limit remain to be
formulated.

\appendix

\section{Conventions, Spinor Decomposition, and Clifford Identities}
\label{app:conventions}

We adopt the Minkowski metric signature $\eta = \mathrm{diag}(-1,1,\dots,1)$. For the Levi-Civita tensor in four dimensions, we define $\epsilon_{0123} = +1$, which implies $\epsilon^{0123} = -1$.

Under the symmetry breaking 
\[
    SO(1,9) \to SO(1,3) \times SO(6)
\]
the ten-dimensional Gamma matrices $\Gamma^M$ decompose into the four-dimensional gamma matrices $\gamma^\mu$ and six-dimensional internal matrices $\rho^a$. Using the 4D chirality matrix $\gamma_5 = i \gamma^0 \gamma^1 \gamma^2 \gamma^3$, the explicit tensor product representation is
\begin{equation}
    \Gamma^\mu = \gamma^\mu \otimes \mathbf{1}_{8} , \quad \Gamma^a = \gamma_5 \otimes \rho^a .
\end{equation}
The ten-dimensional chirality operator decomposes as
\begin{equation}
    \Gamma_{11} = \gamma_5 \otimes \rho_7 ,
\end{equation}
where $\rho_7$ acts as the chirality matrix in the internal space. 

\subsection{Chirality Reduction of $\gamma_5$}

In the following component notation, $\psi^I$ and $\epsilon^I$
denote the four-dimensional Weyl components of fixed chirality
$\chi$; their Majorana-conjugate components are understood
implicitly. The ten-dimensional Weyl condition correlates the
four-dimensional and internal chiralities, and the four-dimensional
components used below satisfy
\begin{equation}
    \gamma_5\psi^I
    =
    \chi\psi^I,
    \qquad
    \gamma_5\epsilon^I
    =
    \chi\epsilon^I,
    \qquad
    \chi=\pm1.
    \label{eq:fixed_four_dimensional_chirality}
\end{equation}

The four-dimensional chirality matrix $\gamma_5$ and the internal
Clifford matrices act on different tensor factors. Therefore,
\begin{equation}
    [\gamma_5,\rho_{a_1\cdots a_n}]=0.
\end{equation}
For any internal Clifford element $\rho_{a_1\cdots a_n}$, it follows
that
\begin{equation}
\begin{split}
    \bar{\epsilon}_2
    \gamma_5\rho_{a_1\cdots a_n}
    \epsilon_1
    &=
    \bar{\epsilon}_2
    \rho_{a_1\cdots a_n}\gamma_5
    \epsilon_1
    \\
    &=
    \chi\,
    \bar{\epsilon}_2
    \rho_{a_1\cdots a_n}
    \epsilon_1 .
\end{split}
\label{eq:gamma5_internal_reduction}
\end{equation}
In particular,
\begin{equation}
    \bar{\epsilon}_I
    (\gamma_5\rho_a)^I{}_J
    \psi^J
    =
    \chi\,
    \bar{\epsilon}_I
    \rho_a{}^I{}_J
    \psi^J .
\end{equation}

We nevertheless retain a single $\gamma_5$ explicitly in the
supersymmetry transformations and commutator bilinears below, so that
the four-dimensional and internal Clifford factors remain manifest.
Equation~\eqref{eq:gamma5_internal_reduction} is used only when the
purely internal closure constraint is extracted. Since
$\chi=\pm1$ is nonzero, this overall chirality factor does not affect
the non-degeneracy argument or the resulting internal
Clifford-algebra constraint.

When two $\gamma_5$ insertions occur in the same spinor chain, they
are simplified using $\gamma_5^2=1$ together with the appropriate
anticommutation relations with the four-dimensional gamma matrices.
For example,
\begin{equation}
    \gamma_5\gamma^\mu\gamma_5=-\gamma^\mu.
\end{equation}
This simplification is distinct from omitting a single $\gamma_5$
insertion.

The ten-dimensional Majorana condition supplies the corresponding
reality relations between the displayed Weyl components and their
Majorana-conjugate components. These reality conditions do not alter
the chirality reduction used above.

\subsection{Useful Clifford-Algebra Identities}
To extract the relevant tensor structures in the commutator evaluations, the following standard Clifford-algebra identities are repeatedly used:
\begin{align}
    \gamma_\mu \gamma_\nu &= \eta_{\mu\nu} + \gamma_{\mu\nu} , \\
    \rho_a \rho_b &= \delta_{ab} + \rho_{ab} , \\
    \rho_a \rho_{bc} &= \rho_{abc} + 2\delta_{a[b}\rho_{c]} , \\
    \rho_{bc} \rho_a &= \rho_{bca} + 2\delta_{a[c}\rho_{b]} .
\end{align}

The bilinears appearing in the effective action, such as $\bar{\epsilon}_I \gamma^\mu \psi^I$ and $\bar{\epsilon}_I (\gamma_5 \rho^a)^I{}_J \psi^J$, are constructed using these orthogonal bases. In particular, $\bar{\epsilon}_I \gamma^\mu \psi^I$ transforms independently as an $SO(1,3)$ vector, while $\bar{\epsilon}_I (\gamma_5 \rho^a)^I{}_J \psi^J$ transforms independently as an $SO(6)$ vector. Since they transform independently under their respective symmetry groups, they define linearly independent kinematic structures in the Ward identity evaluations.

\section{Closure Terms and Coefficient Matching}
\label{app:coefficient_matching}

We first introduce the reference gauge parameter
\begin{equation}
    \Lambda_0
    :=
    v^\mu B_\mu
    +
    v^a\phi_a .
    \label{eq:reference_gauge_parameter}
\end{equation}
Here we choose the normalization
\begin{equation}
\begin{split}
    v^\mu
    &:=
    \bar{\epsilon}_2\gamma^\mu\epsilon_1
    -
    (1\leftrightarrow2),
    \\
    v^a
    &:=
    \bar{\epsilon}_2\gamma_5\rho^a\epsilon_1
    -
    (1\leftrightarrow2).
\end{split}
\label{eq:closure_vector_bilinears}
\end{equation}
We use unit-weight antisymmetrization,
\begin{equation}
    A_{[ab]}
    :=
    \frac{1}{2}
    \left(
        A_{ab}-A_{ba}
    \right).
\end{equation}

The gauge transformations generated by $\Lambda_0$ are
\begin{align}
    \delta_{\Lambda_0}B_\mu
    &=
    -i[B_\mu,\Lambda_0]
    \nonumber\\
    &=
    -v^\nu\mathcal{F}_{\mu\nu}
    -
    v^a\mathcal{F}_{\mu a}
    \nonumber\\
    &=
    v^\nu\mathcal{F}_{\nu\mu}
    +
    v^a\mathcal{F}_{a\mu},
    \label{eq:app_gauge_B}
    \\
    \delta_{\Lambda_0}\phi_a
    &=
    -i[\phi_a,\Lambda_0]
    \nonumber\\
    &=
    -v^\mu\mathcal{F}_{a\mu}
    -
    v^b\mathcal{F}_{ab}
    \nonumber\\
    &=
    v^\mu\mathcal{F}_{\mu a}
    +
    v^b\mathcal{F}_{ba}.
    \label{eq:app_gauge_phi}
\end{align}

Here $\Lambda_0$ fixes the relative tensor structures of the
macroscopic, cross, and internal gauge transformations. The common
overall coefficient generated by the supersymmetry commutator is
incorporated into the actual closure parameter
$\Lambda_{\mathrm{cl}}$ below.

To explicitly see how these gauge variations are generated from the supersymmetry algebra, we evaluate the cross-sector and internal-sector commutators.

Substituting the cross term of the fermion variation,
\begin{equation}
    \delta_{\epsilon_1}^{(1)}\psi
    \supset
    2w_3\mathcal{F}_{\mu a}
    \gamma^\mu\gamma_5\rho^a\epsilon_1,
\end{equation}
into $\delta_{\epsilon_2}^{(1)}B_\nu
=c_1\bar{\epsilon}_2\gamma_\nu\psi$ gives
\begin{equation}
\begin{split}
    [\delta_{\epsilon_1}^{(1)},
     \delta_{\epsilon_2}^{(1)}]B_\nu
    \supset
    2c_1w_3\mathcal{F}_{\mu a}
    \Bigl[
        \bar{\epsilon}_2
        \gamma_\nu\gamma^\mu\gamma_5\rho^a
        \epsilon_1
        -
        (1\leftrightarrow2)
    \Bigr].
\end{split}
\end{equation}
Using
$\gamma_\nu\gamma^\mu
=\delta_\nu^\mu+\gamma_\nu{}^\mu$, its vector part is
\begin{align}
    \left.
    [\delta_{\epsilon_1}^{(1)},
     \delta_{\epsilon_2}^{(1)}]B_\nu
    \right|_{\mathrm{vec}}
    &=
    2c_1w_3\mathcal{F}_{\nu a}
    \Bigl[
        \bar{\epsilon}_2\gamma_5\rho^a\epsilon_1
        -
        (1\leftrightarrow2)
    \Bigr]
    \nonumber\\
    &=
    2c_1w_3v^a\mathcal{F}_{\nu a}
    \nonumber\\
    &=
    -2c_1w_3v^a\mathcal{F}_{a\nu}.
    \label{eq:cross_flux_vector_B}
\end{align}

Substituting the cross-flux term into the internal bosonic transformation $\delta_{\epsilon_2}^{(1)} \phi_a = c_2 \bar{\epsilon}_2 \gamma_5 \rho_a \psi$ gives
\begin{equation}
\begin{split}
    [\delta_{\epsilon_1}^{(1)}, \delta_{\epsilon_2}^{(1)}] \phi_a \supset 2 c_2 w_3 \mathcal{F}_{\mu b} \Bigl[ \bar{\epsilon}_2 \gamma_5 \rho_a \gamma^\mu \gamma_5 \rho^b \epsilon_1 - (1 \leftrightarrow 2) \Bigr] .
\end{split}
\end{equation}
Using $\gamma_5 \gamma^\mu \gamma_5 = -\gamma^\mu$ and $\rho_a \rho^b = \delta_a^{\,b} + \rho_a{}^b$, the vector part is
\begin{equation}
    \left. [\delta_{\epsilon_1}^{(1)}, \delta_{\epsilon_2}^{(1)}] \phi_a \right|_{\mathrm{vec}} = -2 c_2 w_3 v^\mu \mathcal{F}_{\mu a} .
\end{equation}

Similarly, substituting the internal flux term $\delta_{\epsilon_1}^{(1)} \psi \supset w_2 \mathcal{F}_{bc} \rho^{bc} \epsilon_1$ into $\delta_{\epsilon_2}^{(1)} \phi_a$ yields
\begin{equation}
    [\delta_{\epsilon_1}^{(1)}, \delta_{\epsilon_2}^{(1)}] \phi_a \supset c_2 w_2 \mathcal{F}_{bc} \left[ \bar{\epsilon}_2 \gamma_5 \rho_a \rho^{bc} \epsilon_1 - (1 \leftrightarrow 2) \right] .
\end{equation}
Applying the Clifford identity $\rho_a \rho^{bc} = \rho_a{}^{bc} + 2\delta_a^{[b}\rho^{c]}$, this expression naturally splits into two parts. The vector part is $2 c_2 w_2 v^c \mathcal{F}_{ac} = -2 c_2 w_2 v^b \mathcal{F}_{ba}$. The totally antisymmetric part constitutes the remainder:
\begin{equation}
    \Delta \phi_a^{(\mathrm{rem})} = c_2 w_2 \left[ \bar{\epsilon}_2 \gamma_5 \rho_{abc} \epsilon_1 - (1 \leftrightarrow 2) \right] \mathcal{F}^{bc} ,
\end{equation}
which cannot be absorbed into the gauge transformation and provides the exact origin of the Clifford constraint analyzed in Section~3.3.

Summarizing the vector parts of the commutators evaluated above (the purely four-dimensional part $\Delta B_\mu \supset -2 c_1 w_1 v^\nu \mathcal{F}_{\nu\mu}$ is evaluated similarly as shown in Appendix~\ref{app:lorentz_calculation}), we obtain a common overall factor $\mathcal{N} = -2$:
\begin{align}
    \left. [\delta_{\epsilon_1}^{(1)}, \delta_{\epsilon_2}^{(1)}] B_{\mu} \right|_{\mathcal{F}_{\rho\sigma}} &= -2 c_1 w_1 v^{\nu} \mathcal{F}_{\nu\mu} , \\
    \left. [\delta_{\epsilon_1}^{(1)}, \delta_{\epsilon_2}^{(1)}] B_{\mu} \right|_{\mathcal{F}_{\rho a}} &= -2 c_1 w_3 v^{a} \mathcal{F}_{a\mu} , \\
    \left. [\delta_{\epsilon_1}^{(1)}, \delta_{\epsilon_2}^{(1)}] \phi_{a} \right|_{\mathcal{F}_{\mu b}} &= -2 c_2 w_3 v^{\mu} \mathcal{F}_{\mu a} , \\
    \left. [\delta_{\epsilon_1}^{(1)}, \delta_{\epsilon_2}^{(1)}] \phi_{a} \right|_{\mathcal{F}_{bc}} &= -2 c_2 w_2 v^{b} \mathcal{F}_{ba} .
\end{align}
Comparison with
Eqs.~\eqref{eq:app_gauge_B} and
\eqref{eq:app_gauge_phi} shows that closure into a single gauge
transformation requires the four coefficient products to coincide:
\begin{equation}
    c_1w_1
    =
    c_1w_3
    =
    c_2w_3
    =
    c_2w_2
    \equiv
    \kappa_{\mathrm{cl}} .
    \label{eq:closure_coefficient_matching}
\end{equation}
The symbol $\kappa_{\mathrm{cl}}$ denotes the common closure
coefficient and should not be confused with the deformation scale
conventionally denoted by $\kappa$ in the
$\kappa$-Minkowski literature.

Since a gauge transformation is linear in its parameter, the common
overall factor in the supersymmetry commutator can be absorbed into
the actual closure parameter. With the present commutator convention,
we define
\begin{equation}
\begin{split}
    \Lambda_{\mathrm{cl}}
    &:=
    -2\kappa_{\mathrm{cl}}\Lambda_0
    \\
    &=
    -2\kappa_{\mathrm{cl}}
    \left(
        v^\mu B_\mu
        +
        v^a\phi_a
    \right).
\end{split}
\label{eq:actual_closure_gauge_parameter}
\end{equation}
The complete vector parts of the two bosonic commutators can then be
written as
\begin{align}
    \left.
    [\delta_{\epsilon_1}^{(1)},
     \delta_{\epsilon_2}^{(1)}]
    B_\mu
    \right|_{\mathrm{vec}}
    &=
    -2\kappa_{\mathrm{cl}}\,
    \delta_{\Lambda_0}B_\mu
    =
    \delta_{\Lambda_{\mathrm{cl}}}B_\mu,
    \label{eq:closure_as_gauge_B}
    \\
    \left.
    [\delta_{\epsilon_1}^{(1)},
     \delta_{\epsilon_2}^{(1)}]
    \phi_a
    \right|_{\mathrm{vec}}
    &=
    -2\kappa_{\mathrm{cl}}\,
    \delta_{\Lambda_0}\phi_a
    =
    \delta_{\Lambda_{\mathrm{cl}}}\phi_a.
    \label{eq:closure_as_gauge_phi}
\end{align}
Thus, Eq.~\eqref{eq:closure_coefficient_matching} fixes the relative
coefficients of the four gauge-transformation structures, whereas
the common overall normalization is carried by
$\Lambda_{\mathrm{cl}}$.

In particular,
Eq.~\eqref{eq:closure_coefficient_matching} contains
\begin{equation}
    c_1w_3=c_2w_3.
\end{equation}
On the generic cross-coupled branch $w_3\neq0$, this gives
\begin{equation}
    c_1=c_2.
\end{equation}

\section{Four-Dimensional Remainder and Lorentz Absorption}
\label{app:lorentz_calculation}

The purely four-dimensional commutator contains both the standard
gauge-transformation structure and a distinct dual-flux remainder.
Substituting the four-dimensional fermion variation
\begin{equation}
    \delta_{\epsilon_1}^{(1)}\psi
    \supset
    w_1\mathcal{F}_{\rho\sigma}
    \gamma^{\rho\sigma}\epsilon_1
\end{equation}
into
\begin{equation}
    \delta_{\epsilon_2}^{(1)}B_\mu
    =
    c_1\bar{\epsilon}_2\gamma_\mu\psi
\end{equation}
gives
\begin{equation}
\begin{split}
    \left.
    [\delta_{\epsilon_1}^{(1)},
     \delta_{\epsilon_2}^{(1)}]
    B_\mu
    \right|_{\mathcal{F}_{\rho\sigma}}
    =
    c_1w_1\mathcal{F}_{\rho\sigma}
    \Bigl[
        \bar{\epsilon}_2
        \gamma_\mu\gamma^{\rho\sigma}
        \epsilon_1
        -
        (1\leftrightarrow2)
    \Bigr].
\end{split}
\label{eq:four_dimensional_commutator}
\end{equation}

Using
\begin{equation}
    \gamma_\mu\gamma^{\rho\sigma}
    =
    \gamma_\mu{}^{\rho\sigma}
    +
    2\delta_\mu^{[\rho}\gamma^{\sigma]},
\label{eq:four_dimensional_gamma_split}
\end{equation}
we separate the vector part from the totally antisymmetric
three-gamma part.

The vector contribution is
\begin{align}
    \left.
    [\delta_{\epsilon_1}^{(1)},
     \delta_{\epsilon_2}^{(1)}]
    B_\mu
    \right|_{\mathrm{vec},\,\mathcal{F}_{\rho\sigma}}
    &=
    2c_1w_1
    \mathcal{F}_{\rho\sigma}
    \delta_\mu^{[\rho}v^{\sigma]}
    \nonumber\\
    &=
    2c_1w_1
    \mathcal{F}_{\mu\nu}v^\nu
    \nonumber\\
    &=
    -2c_1w_1
    v^\nu\mathcal{F}_{\nu\mu}.
\label{eq:four_dimensional_vector_part}
\end{align}
Here we used the definition of \(v^\mu\) in
Eq.~\eqref{eq:closure_vector_bilinears} and the antisymmetry
\(\mathcal{F}_{\mu\nu}=-\mathcal{F}_{\nu\mu}\).
Equation~\eqref{eq:four_dimensional_vector_part} is precisely the
four-dimensional gauge-transformation structure appearing in
Eq.~\eqref{eq:app_gauge_B}, with the common overall factor
\(-2c_1w_1\).

The remaining three-gamma contribution is
\begin{equation}
\begin{split}
    \Delta B_\mu^{(\mathrm{rem})}
    :=
    c_1w_1\mathcal{F}_{\rho\sigma}
    \Bigl[
        \bar{\epsilon}_2
        \gamma_\mu{}^{\rho\sigma}
        \epsilon_1
        -
        (1\leftrightarrow2)
    \Bigr].
\end{split}
\label{eq:four_dimensional_three_gamma_remainder}
\end{equation}
Since
\begin{equation}
    \mathcal{F}_{\rho\sigma}
    \gamma_\mu{}^{\rho\sigma}
    =
    \mathcal{F}^{\rho\sigma}
    \gamma_{\mu\rho\sigma},
\end{equation}
we use the four-dimensional identity
\begin{equation}
    \gamma_{\mu\rho\sigma}
    =
    -i\epsilon_{\mu\rho\sigma\nu}
    \gamma^\nu\gamma_5 .
\label{eq:four_dimensional_three_gamma_duality}
\end{equation}
It follows that
\begin{align}
    \Delta B_\mu^{(\mathrm{rem})}
    &=
    -ic_1w_1
    \epsilon_{\mu\rho\sigma\nu}
    \mathcal{F}^{\rho\sigma}
    \Bigl[
        \bar{\epsilon}_2
        \gamma^\nu\gamma_5
        \epsilon_1
        -
        (1\leftrightarrow2)
    \Bigr]
    \nonumber\\
    &=
    -i\chi c_1w_1
    \epsilon_{\mu\rho\sigma\nu}
    \mathcal{F}^{\rho\sigma}
    v^\nu ,
\label{eq:four_dimensional_remainder_before_dual}
\end{align}
where the fixed-chirality relation
\(\gamma_5\epsilon_i=\chi\epsilon_i\), with \(\chi=\pm1\), has been
used.

Defining the dual four-dimensional flux by
\begin{equation}
    \widetilde{\mathcal{F}}_{\nu\mu}
    :=
    \frac{1}{2}
    \epsilon_{\nu\mu\rho\sigma}
    \mathcal{F}^{\rho\sigma},
\label{eq:dual_four_dimensional_flux}
\end{equation}
and using
\begin{equation}
    \epsilon_{\mu\rho\sigma\nu}
    \mathcal{F}^{\rho\sigma}
    =
    -\epsilon_{\nu\mu\rho\sigma}
    \mathcal{F}^{\rho\sigma}
    =
    -2\widetilde{\mathcal{F}}_{\nu\mu},
\end{equation}
we obtain
\begin{equation}
    \Delta B_\mu^{(\mathrm{rem})}
    =
    2i\chi c_1w_1
    v^\nu
    \widetilde{\mathcal{F}}_{\nu\mu}.
\label{eq:four_dimensional_dual_remainder}
\end{equation}

Thus, the complete four-dimensional flux contribution to the
commutator is
\begin{equation}
\begin{split}
    \left.
    [\delta_{\epsilon_1}^{(1)},
     \delta_{\epsilon_2}^{(1)}]
    B_\mu
    \right|_{\mathcal{F}_{\rho\sigma}}
    =
    -2c_1w_1
    v^\nu\mathcal{F}_{\nu\mu}
    +
    2i\chi c_1w_1
    v^\nu\widetilde{\mathcal{F}}_{\nu\mu}.
\end{split}
\label{eq:complete_four_dimensional_flux_commutator}
\end{equation}
The first term is the standard gauge-transformation contribution,
whereas the second term is the distinct four-dimensional dual-flux
remainder studied in Section~\ref{sec:lorentz_absorption}.

On the nontrivial four-dimensional branch $c_1w_1\neq0$, define
\begin{equation}
    \mathcal{C}_4
    :=
    2i\chi c_1w_1.
\end{equation}
Exact absorption of the remainder by the Lorentz-type variation is the
condition
\begin{equation}
    \mathcal{C}_4\,
    \widetilde{\mathcal{F}}_{\nu\mu}
    =
    M_{\nu\mu\lambda}B^\lambda,
\end{equation}
in agreement with Eq.~\eqref{eq:exact_linear_absorption}. Substituting
Eq.~\eqref{eq:M_hodge_dual_exact} and using
$\mathcal{C}_4\neq0$ gives
\begin{equation}
    \frac{1}{2}
    \epsilon_{\nu\mu\rho\sigma}
    \mathcal{F}^{\rho\sigma}
    =
    -\Theta\,
    \epsilon_{\tau\nu\mu\lambda}
    m^\tau B^\lambda.
    \label{eq:exact_absorption_epsilon_form}
\end{equation}
Multiplying Eq.~\eqref{eq:exact_absorption_epsilon_form} by
$\epsilon^{\alpha\beta\nu\mu}$ and using
\begin{equation}
    \epsilon^{\alpha\beta\nu\mu}
    \epsilon_{\nu\mu\rho\sigma}
    =
    -2
    \left(
        \delta^\alpha_\rho\delta^\beta_\sigma
        -
        \delta^\alpha_\sigma\delta^\beta_\rho
    \right),
\end{equation}
together with the corresponding contraction on the right-hand side,
we obtain
\begin{equation}
    -2\mathcal{F}^{\alpha\beta}
    =
    2\Theta
    \left(
        m^\alpha B^\beta
        -
        m^\beta B^\alpha
    \right).
\end{equation}
Therefore,
\begin{equation}
    \mathcal{F}^{\alpha\beta}
    =
    -\Theta
    \left(
        m^\alpha B^\beta
        -
        m^\beta B^\alpha
    \right),
\end{equation}
which is Eq.~\eqref{eq:exact_flux_relation}. Using
$\mathcal{F}^{\mu\nu}=i[B^\mu,B^\nu]$ then gives
\begin{equation}
    [B^\mu,B^\nu]
    =
    i\Theta
    \left(
        m^\mu B^\nu
        -
        m^\nu B^\mu
    \right).
\end{equation}

\section*{Declaration of generative AI and AI-assisted technologies in the manuscript preparation process}

During the preparation of this work, the author used generative AI tools, including Gemini and ChatGPT, in order to improve the readability, check for typographical errors, and refine the English expression of the manuscript. After using these services, the author reviewed and edited the content as needed and takes full responsibility for the content of the published article.

\end{document}